\documentclass[aps,twocolumn,showpacs,prl]{revtex4}
\usepackage{graphicx} 
\usepackage{amsmath}
\usepackage{times}
\begin{document}
\title{Evolutionary games and the emergence of complex networks}
\author{Holger Ebel}
\email{ebel@theo-physik.uni-kiel.de}
\affiliation{Institute for Theoretical Physics, University of Kiel,
Leibnizstr.\ 15, D-24098 Kiel, Germany}

\author{Stefan Bornholdt}
\email{bornholdt@izbi.uni-leipzig.de}
\affiliation{Interdisciplinary Center for Bioinformatics, 
University of Leipzig, Kreuzstr.\ 7b, D-04103 Leipzig, Germany} 
 
\begin{abstract} 
The emergence of complex networks from evolutionary games 
is studied occurring when agents are allowed to switch interaction partners. 
For this purpose a coevolutionary iterated Prisoner's Dilemma game is defined on 
a random network with agents as nodes and games along the links. The  
agents change their neighborhoods to improve their payoff. 
The system relaxes to stationary states  
corresponding to cooperative Nash equilibria with the additional 
property that no agent can improve its payoff by changing its neighborhood.
Small perturbations of the system lead to avalanches of strategy readjustments  
reestablishing equilibrium. As a result of the dynamics, the network of 
interactions develops non-trivial topological properties as a broad degree 
distribution suggesting scale-free behavior, small-world characteristics,
and assortative mixing. 
\end{abstract}
\maketitle

Economic relationships evolve: If the fish you bought on the fish market 
is spoiled you will probably switch the dealer next time. On larger scales,  
markets may develop into complex interaction structures solely from agents 
using their freedom of choosing their partners. 

Statistical physics has a tradition of studying systems of interacting 
agents motivated by economic or social interactions. 
One example is evolutionary game theory on the lattice 
where agents can be thought of as particles with 
generalized interactions and internal memory. These properties give rise to
complex spatiotemporal dynamics when agents try to locally optimize their payoffs 
\cite{nowak/may:1992,herz:1994,lindgren/nordahl:1994}.
The goal of this paper is to relax a further constraint in this class of 
models, namely the fixed interaction topology. Agents may use the choice 
of interaction partners as another means to locally optimize their payoff 
and, thereby, shape the interaction matrix and the topology of the system. 

An interesting aspect of agent systems with evolving neighborhoods is 
the structure of the emerging interaction networks. The theoretical 
physics community has focused on complex network structures in diverse 
natural systems recently \cite{strogatz:2001,albert/barabasi:2002}.
Game theoretic models on evolving networks may provide a fresh view 
on the emergence of complex networks in systems where
strategic interactions occur. One example are economic systems 
where complex global structures and dynamics are observed 
\cite{arthur:1999,kirman:2002a}.
 
The dynamics of games on static networks has been studied
recently, namely the question of stability of cooperative behavior in 
the iterated Prisoner's Dilemma (IPD)
\cite{abramson/kuperman:2001,kim/choi:2002,ebel/bornholdt:2002b}, 
also under the influence of noise in the spatial relationships of agents 
\cite{cohen/axelrod:2001}. The question how the game itself may 
affect network topology was asked by Zimmermann et al.\ in a 
simple setting of cooperating and defecting players, leading to a 
network that self-organizes into a steady state with highly 
connected cooperators \cite{zimmermann/miguel:2001}. 

In the following, let us consider the IPD  with memory
on an evolving network. Details of this game on static networks
can be found in Ref.\ \cite{ebel/bornholdt:2002b}.  
The Prisoner's Dilemma  \cite{axelrod/hamilton:1981,axelrod:1984} is a
two-person game defined
by the payoff matrix
\begin{equation}
A=
\begin{pmatrix}
3 & 0 \\
5 & 1 
\end{pmatrix}=(a_{ij})_{i,j \, \in \, \{1,2\}},\label{payoff-matrix}
\end{equation}
with the entries $a_{ij}$ being the payoffs of agent 1 playing strategy 
$s_i$ against agent 2 using strategy $s_j$ ($s_1$, cooperating; $s_2$, defecting). 
In general, the entries $a_{ij}$ have to fulfill 
$a_{12} < a_{22} < a_{11} < a_{21}$ and $a_{12}+a_{21} < 2 a_{11}$.
\begin{table}\centering
\begin{ruledtabular}
\begin{tabular}{cc}
History & Action \\
\hline 
0 & 1 \\
1 & 1 \\
First move & 1 \\
\end{tabular}
\end{ruledtabular}
\caption{\label{lookup-table} Representation of an agent's strategy 
(0, defection; 1, cooperation). A player with the above 
strategy cooperates, no matter whether its opponent has cooperated or 
defected in the last move.}
\end{table}

In the IPD, a strategy is interpreted as a map of an agent's 
knowledge to an action. Here, knowledge is defined by the number of the $m$
previous moves the agent can remember. We 
will confine the strategy space to strategies with $m=1$, 
i.e., one agent knows only about the latest move of its opponent. 
At the beginning of an encounter, there is no previous move. Therefore, 
the first action is also part of the strategy (Table \ref{lookup-table}). 
It can be viewed as a lookup table where each history is mapped 
to an action. In the case of $m=1$, there are $2^{2^{m}+1}=8$ possible 
strategies. The strategies ``always defect" (000) and ``tit for tat" 
(011) are always Nash equilibria of this finite normal form game.
Initially, a random network of a given average degree 
(i.e., number of a node's next neighbors) $\langle k \rangle$ is generated 
resulting in a Poissonian degree distribution.
Strategies are assigned to the players at random. The coevolutionary 
dynamics consists of one part for the evolution of the strategies and 
another for network evolution. In each iteration cycle the following 
steps take place. 
{\em Strategy evolution}:  
(i) One agent $i$ is chosen randomly and its strategy is mutated to 
a strategy picked at random. 
(ii) The mutated agent plays against its neighbors and its payoff 
is compared to its payoff before the mutation. In case of a payoff increase 
the mutation is accepted and the payoffs of all neighbors are updated. 
This strategy update has been first used in Ref.\ \cite{ebel/bornholdt:2002b}. 
Step (ii) corresponds to the assumptions that changing the strategy 
may bear some costs (risk) for the player and that mutations occur on 
a time scale slower than the time scale of the game. 
{\em Network evolution}: 
(i) With probability $\alpha$, one randomly chosen agent $i$ is connected 
to a new neighbor taken at random. 
(ii) If the new connection leads to an increase in average 
payoff, player $i$ will 
accept the new connection and remove the link to the neighbor it scores 
worst against. In case player $i$ had no neighbor before, 
it accepts the connection, and a link between a random pair of players 
will be removed to keep the average degree $\langle k \rangle$ constant 
\footnote{The case of growing $\langle k \rangle$ 
only adds a drift $\langle k \rangle_t \propto \sqrt{t}$ 
to the dynamics. This is caused by the occurrence of disconnected nodes 
with probability $P_t(k=0) \propto 1 / \sqrt{t}$. The resulting clustering 
$C_{\Delta,t} \propto \sqrt{t}$ is fully reproduced by (\ref{eq_c}).}.
Thereafter, all payoffs are updated. 
Note that the information used by an agent (its own payoff) is fully local. 

The iteration of these steps eventually leads to stationary states.
The choice of the 
value of $\alpha$ that determines the speed of network evolution is 
not critical and convergence has been verified for values between 
$\alpha=0.001$ and $1$. The stationary states, where no further 
increase in payoff is possible for any agent, are cooperative and 
dominated by the strategy ``tit for tat'' (011). 
With regard to strategies, the stationary states correspond to 
game theoretic Nash equilibria \cite{nash:1950}. 
Network evolution here does not interfere with the notion of Nash 
equilibria. In a stationary state no player can improve 
its payoff, neither by changing its strategy nor by choosing a different 
neighborhood. In the remainder of this paper, such an equilibrium will 
be called a {\em network Nash equilibrium}.

\begin{figure}
\centering
\includegraphics[width=5.5cm,angle=-90]{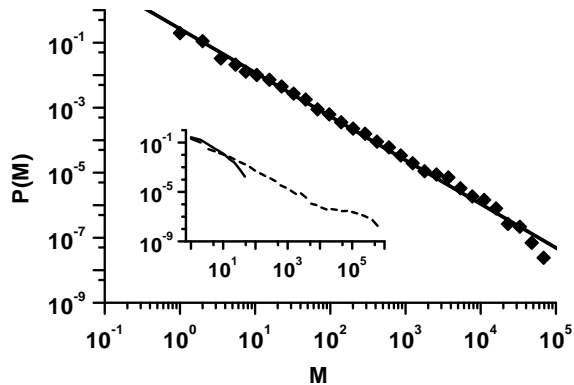}
\caption{Distribution $P(M)$ of avalanche size $M$ in the critical regime. 
$P(M)$ exhibits scale-free behavior with a scaling exponent of 
$\gamma =-1.35\pm0.04$. The inset shows the sub- and supercritical regimes 
($N=500$, $\langle k \rangle=5$, $a_{21}=3.5, 5.0, 5.1$).}
\label{fig1}
\end{figure}
One essential property of this spatially extended coevolutionary game
is the stability of the network Nash equilibria. 
Exploring the IPD on static networks \cite{ebel/bornholdt:2002b}, 
perturbations of the game's Nash equilibria have been observed to cause
avalanches of $M$ mutation events until stationary states are reestablished. 
Depending on the crucial parameter of the payoff matrix, the temptation 
to defect $a_{21}$, three different regimes of avalanche dynamics were
found. First, for low temptations, the distribution of 
avalanche size $P(M)$ is subcritical with few mutation events 
necessary to reach again an equilibrium. In an intermediate range 
of $a_{21}$, scale-free behavior occurs as the distribution of 
avalanche size exhibits a power law with exponent $\gamma=-1.39\pm0.10$. 
For high values of $a_{21}$, the game becomes supercritical with a 
probability for very large avalanches much higher compared to the 
critical regime. 

Let us similarly perturb the network Nash equilibrium by randomly 
changing the strategy of an arbitrary player \footnote{One can 
additionally change a link which, however, does not significantly 
alter the system's response to the perturbation.}. 
The deviating strategy offers new opportunities for strategy changes 
to the neighboring players. The perturbation then can spread which 
leads to an avalanche of mutation events. Additionally, payoff can 
be improved by changing one's neighborhood which results in a 
network evolution closely connected to the strategy evolution of the 
individual agents.

One finds that the model with network dynamics considered here results 
in a similar avalanche dynamics as observed for the static network case. 
For temptations to defect $a_{21} \leq 4.8$ and $a_{21} \geq 5.2$, 
the subcritical and supercritical regimes emerge 
(inset in Fig.\ \ref{fig1}). The critical regime exists in the same range 
$4.9 \leq a_{21} \leq 5.1$ as for static networks with same mean degree 
(Fig.\ \ref{fig1}). In addition, the scaling exponent $\gamma = -1.35\pm0.04$ 
agrees with the static case. Surprisingly, however, evolution of 
network structure is profoundly affected by the avalanche dynamics.

\begin{figure}
\centering
\includegraphics[width=5.5cm,angle=-90]{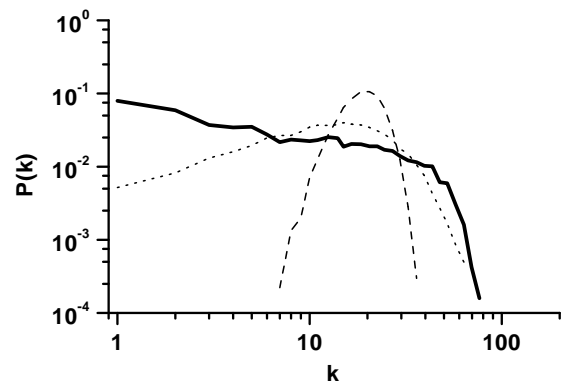}
\caption{Degree distributions $P(k)$ for critical (solid line, $a_{21}=5.0$)
and subcritical (dotted line, $a_{21}=3.5$) avalanche dynamics. The dashed
plot shows the initial Poissonian degree distribution. ($N=500$, $\langle k 
\rangle =20$).} 
\label{fig2}
\end{figure}
After an avalanche, network evolution comes to an end when a 
network Nash equilibrium is reached. The equilibrium network, which emerges
in the course of subsequent perturbations and avalanches, 
strongly differs from the initial random graph. 
For practical purposes we here speak of an equilibrium network 
when the first three cumulants of the degree distribution do 
not change over a period of time much larger than the time needed 
to reach these values.
The first three cumulants 
are given by the mean degree $\langle k \rangle$ (being constant 
because of step (ii) of the network evolution process), the variance 
$\sigma^2$ and the {\em skewness}
$\lambda_3 = \langle (k-\langle k \rangle)^3 \rangle / \sigma^3$.
The stationary values of the cumulants for different mean degrees 
and the three distinct regimes of avalanche dynamics are given in 
Table \ref{tab_pk}. Independent of mean degree and temptation to defect, 
one observes a strong broadening of the degree distribution, 
up to an increase of the variance $\sigma^2$ by a factor of 20 
compared to the initial random network with Poissonian degree 
distribution ($\sigma^2_0$). The skewness $\lambda_3$ is significantly 
higher than expected for the Poissonian distribution ($\lambda_3^0$) 
indicating a more asymmetric distribution with a longer tail to the 
right. The dependence of the degree distribution on the different 
regimes of avalanche dynamics can be seen in Fig.\ \ref{fig2} 
where the degree distributions for critical and subcritical 
avalanche dynamics are plotted. In the critical regime 
(and similarly in the supercritical regime), the broad degree 
distribution resembles a power law. In the range of 
$\langle k \rangle$ where critical dynamics are possible
\cite{ebel/bornholdt:2002b} 
the cutoff of the degree distribution grows with increasing 
$\langle k \rangle$, whereas the degree distribution is independent 
of the number of agents $N$.
\begin{table}
\begin{ruledtabular}
\begin{tabular}{cccccc}
$\langle k \rangle$ & $a_{21}$ & $\sigma^2$ & $\sigma^2_0$
& $\lambda_3$ & $\lambda_3^0$ \\ \hline 
5 & 3.5 & 10.4(4) & 5 & 0.964(74) & 0.447 \\ 
5 & 5.0 & 19.3(12) & 5 & 1.22(13) &  0.447  \\ 
5 & 5.1 & 20.6(12) & 5 & 1.39(13) & 0.447 \\ 
20 & 3.5 & 125(7) & 20 & 0.873(89) & 0.224 \\ 
20 & 5.0 & 261(8) & 20 & 0.753(72) & 0.224  \\ 
20 & 5.4 & 398(13) & 20 & 0.906(36) & 0.224 \\ 
\end{tabular}
\caption{Cumulants of the degree distribution $P(k)$ 
($N=500$, errors are given by standard deviations).}
\label{tab_pk}
\end{ruledtabular}
\end{table}
Note that although the quantity selected for is the sum 
of the payoffs an agent receives, it does not select for 
high number of neighbors as the number of neighbors is kept 
constant for the agent who decides. However, the agent who 
receives the link appears to be more or less attractive 
depending on its strategy and this reflects in the emerging
link distribution. 

\begin{table}
\begin{ruledtabular}
\begin{tabular}{ccccccccc}
$\langle k \rangle$ & $a_{21}$ & $C_\Delta$ & $C'$ & $C_0$ &
 $\ell$ & $\ell'$ & $\ell_0$ & $r$ \\ \hline 
5 & 3.5 & 0.0164(21) & 0.014 & 0.010 & 3.98(3) & 3.9 & 3.55 & 0.047(22) \\ 
5 & 5.0 & 0.0298(36) & 0.025 & 0.010 & 3.95(4) & 3.2 & 3.55 & 0.115(25)\\ 
5 & 5.1 & 0.0329(39) & 0.026 & 0.010 & 3.95(5) & 3.2 & 3.55 & 0.125(23)\\ 
20 & 3.5 & 0.0651(25) & 0.064 & 0.040 & 2.41(1) & 2.0 & 2.07 & 0.006(11)\\ 
20 & 5.0 & 0.106(2) & 0.10 & 0.040 & 2.55(2) & 1.9 & 2.07 & 0.019(14)\\ 
20 & 5.4 & 0.162(7) & 0.15 & 0.040 & 2.69(4) & 1.9 & 2.07 & 0.064(18)\\ 
\end{tabular}
\caption{Clustering coefficient $C_\Delta$, average shortest 
path length $\ell$, and Pearson coefficient $r$ of the evolved 
networks ($N=500$, errors are given by standard deviations). }
\label{tab_sw}
\end{ruledtabular}
\end{table}
Besides the degree distribution, important properties when exploring a 
complex network's topology are also clustering and average shortest path length. 
The clustering coefficient $C_\Delta$ is a measure of the probability 
that two nodes share a common neighbor \cite{newman/watts:2001}
\begin{equation}
C_\Delta = \frac{3 \times (\text{number of fully connected triples})}{\text{number
of triples}}.
\label{eq_cn}
\end{equation}
$C_\Delta$ is the fraction of fully connected triples, with a 
triple being a connected subgraph which contains exactly three nodes. 
These values are compared to the clustering coefficient of a Poissonian 
random network $C_0$ and the clustering coefficient $C'$ we derived for a 
random network with identical degree distribution but randomly assigned 
links \cite{davidsen/bornholdt:2002}
\begin{equation}
C' = \frac{1}{\langle k \rangle N}
\left (\frac{\langle k^2 \rangle}{\langle k \rangle}-1\right)^2.
\label{eq_c}
\end{equation}

Comparison of the experimental clustering $C_\Delta$ with $C_0$ shows 
an enhanced clustering (Table \ref{tab_sw}).
Since $C_\Delta$ is reproduced 
well by Eq.\ (\ref{eq_c}) the increase in clustering can be entirely 
traced to the change of the degree distribution.

The average $\ell$ of the shortest path lengths between two nodes is
determined for the giant component of 
the network always containing more than 95\%
of all nodes. 
Similarly to the treatment of the clustering coefficient, shortest 
path lengths for Poissonian random networks ($\ell_0$) and for 
random networks with identical degree distributions ($\ell'$) 
are calculated \cite{davidsen/bornholdt:2002}.
The observed values of $\ell$ are slightly larger than the average 
shortest paths in a Poissonian random network resulting in still 
very short paths between any two nodes in the network (Table 
\ref{tab_sw}).
The estimate $\ell'$ does not yield results as 
good as $C'$ in the case of the clustering coefficient. 
Obviously, for the shortest path length $\ell$, the assumption of 
randomly assigned links \cite{newman/watts:2001,davidsen/bornholdt:2002}
is not sufficient for an accurate estimate.

To gain further insight into the structure of the evolved networks, 
the Pearson coefficient $r$ is calculated. It provides a 
measure of the degree correlation between neighboring nodes 
\cite{newman:2002c} 
\begin{equation}  
r = \frac{1}{\sigma_Q^{2}}  \sum_{\mu,\nu} \mu \nu \biggl(
Q(\mu,\nu) - Q(\mu) Q(\nu) \biggr).
\end{equation}
$Q(\mu)$ is the distribution of the {\em remaining degree} 
$\mu$ (i.e., $\mu = k -1$) and $Q(\mu,\nu)$ the joint probability 
distribution that the two nodes at the ends of a randomly 
chosen link have the remaining degrees $\mu$ and $\nu$. 
A positive Pearson coefficient indicates that nodes with a 
high degree prefer linking to other highly connected nodes
({\em assortative mixing}).
For $r < 0$ high degree nodes tend to link to poorly 
connected agents ({\em disassortative mixing}).
No preference is given when such correlations are 
absent, $r = 0$. The evolved networks studied here 
are neutral in the subcritical avalanche regime, as expected 
for the initial Poissonian random network (Table \ref{tab_sw}).
However, for critical and supercritical behavior, assortative 
mixing is observed (comparable in size to the assortative mixing 
of real-world graphs as, for example, co-authorship 
networks \cite{newman:2002c}). 

To summarize we have studied a model for a coevolutionary 
Prisoner's Dilemma game on an evolving network
where each agent is allowed to alter its neighborhood 
in order to enhance its payoff, thereby shaping the topology 
of the network. 
One observes that the network, after an initial perturbation
and a subsequent avalanche of rearrangements, relaxes into a
stationary state which is stable w.r.t.\ strategy variations 
(Nash equilibrium), as well as against local topology 
changes (network Nash equilibrium). In this equilibrium, even 
a change of topology cannot enhance any player's payoff. 
While the dynamics of the relaxation process is similar 
to the corresponding model on static networks  
without topology updates, the equilibrium structure of the 
network, emerged in the course of avalanches, strongly 
deviates from the initial random graph. 
The network evolves to a statistically stationary state with a 
broad degree distribution, suggesting scale-free behavior 
in the critical avalanche regime. It exhibits small-world behavior 
\cite{watts/strogatz:1998}, and in the critical and supercritical 
regimes also shows assortative mixing, a property of social networks 
not shared by many standard models \cite{newman:2002c}.

The spatial coevolutionary game studied here is a minimal non-trivial
model of network evolution driven by game theoretic interactions.
It may help in understanding the emergence of global organization from local
interactions as observed in social and economic systems.
It also provides an alternative mechanism for the evolution of complex networks 
possibly relevant for real-world systems where simple 
growth models as, for example, {\em preferential linking}
\cite{albert/barabasi:2002}, are not easily applicable.

\begin{acknowledgements}
We thank R.\ Axelrod, P.\ Holme, and K.\ Klemm for useful 
discussions and comments. This work has been partly supported 
by the Deutsche Forschungsgemeinschaft (German Science Foundation). 
H.~E.\ gratefully acknowledges support by  the Studienstiftung des 
deutschen Volkes (German National Merit Foundation). 
\end{acknowledgements}

\end{document}